\documentclass[12pt,comp]{article}
\usepackage[T2A]{fontenc}
\usepackage[cp1251]{inputenc}
\usepackage[english]{babel}
\usepackage{graphicx,caption,subcaption}
\usepackage{amsmath}
\usepackage{amssymb}
\usepackage{ulem}
\usepackage{makeidx}
\usepackage{fancyhdr}

\begin{document}

\begin{center} \noindent {\textbf{\Large   Finding the
effective structure parameters for \\suspensions of nano-sized
insulating particles from low-frequency impedance measurements}}
 \end{center}
\bigskip
\begin{center}
{$\rm M. Ya. Sushko^{\it a}$\footnote{$\rm Corresponding$ author,
e-mail: mrs@onu.edu.ua}}, V. Ya. $\rm Gotsulskiy^{\it b}$,  and M.
V. $\rm Stiranets^{\it b}$

$\rm ^{\it b}Department$ of Theoretical Physics, Mechnikov
National University

$\rm  ^{\it a}Department$ of General and Chemical Physics,
Mechnikov National University

2 Dvoryanska St., Odesa 65026, Ukraine \end{center}

\begin{abstract}
Based on the method of compact groups of inhomogeneities, we
formulate new mixing rules for suspensions of charged insulating
particles. They express the quasistatic conductivity and
permittivity of a suspension in terms of the effective geometric
and dielectric parameters of the particles, electric double layers
(EDLs), and suspending liquid. Also, we present our low-frequency
impedance measurements of the conductivity and permittivity of
$\rm {Al_2O_3}$-isopropyl alcohol nanofluids as functions of $\rm
{Al_2O_3}$-particle volume concentration. Our rules give good fits
for most of these data and allow us to estimate, among other
things, the effective thickness, conductivity, and permittivity of
the EDLs. The experimentally-recovered values agree well with
elementary theoretical estimates suggesting the charging of the
particles through preferential adsorption of contaminant ions. The
possible effects of other mechanisms on the effective conductivity
and permittivity of suspensions are also discussed.
\end{abstract}

\vspace{2pc} \noindent{\it Keywords}: Electrical conductivity,
Permittivity, Suspension, Nanofluid, Electric double layer,
Compact group method, Impedance spectroscopy


\section{Introduction}

The nature and structure of  interphase layers in dispersions and
nanofluids (liquid-based suspensions of nano-sized particles) have
been under intensive study in the context of their effect on
various bulk properties of the entire system. In particular, it is
recognized that the diffuse electrical double layers (EDLs) play
an essential role in the formation of the conductivity
$\sigma_{\rm eff}$ and permittivity $\varepsilon_{\rm eff}$ of
nanofluids
\cite{bib:Dukhin74,bib:Hunter81,bib:Lyklema95,bib:Hunter01,bib:Morgan03,bib:Delgado05,bib:Ohshima12}.
However, the details of pertinent mechanisms and their
incorporating into a consistent quantitative theory of the
effective electrical properties of nanofluids still remain to be
investigated. For example, the classical Maxwell--Garnett
\cite{bib:Maxwell1873,bib:Garnett04} and Bruggemann
\cite{bib:Bruggeman35} mixing rules fail to explain a dramatic
enhancement of $\sigma_{\rm eff}$ upon addition of small amounts
of insulating nanoparticles. The reasons are that  they, first,
ignore the actual microstructure of the system and, second, are
one-particle approximations. Modern approaches (see
\cite{bib:Dukhin74,bib:Hunter81,bib:Lyklema95,bib:Hunter01,bib:Morgan03,bib:Delgado05,bib:Ohshima12})
to the problem represent various sophisticated modifications of
the electrokinetic theories
\cite{bib:OBrien78,bib:Saville79,bib:OBrien81,bib:DeLacey81} and
deal with a set of coupled differential equations for the flow
field and for quantities related to the ion density and electric
field distributions around a hard particle. The results heavily
depend on model assumptions about the factors and mechanisms
responsible for the formation of these distributions and are
usually restricted to situations where the total volume
concentration of the particles and their EDLs is very small. The
latter means that the electromagnetic interactions between the
structural inhomogeneities, formed by the particles and their
EDLs, are weak, so that finding the effective conductivity and
permittivity of the entire suspension is actually reduced to
finding those of a single inhomogeneity and then corrections to
them in the lowest orders of various perturbation techniques. Most
known attempts at analyzing the effective parameters of
concentrated suspensions are based upon Happel's
\cite{bib:Happel58} and Kuwabara's \cite{bib:Kuwabara59} cell
models. On the outer surface of the unit cell, the governing
electrokinetic equations are subject to certain boundary
conditions of the Neumann or Dirichlet type. Being supposed to
account for the averaged effects of the surrounding medium on the
cell, the boundary conditions are constructed by applying
mean-field considerations to the local fields (see
\cite{bib:Levine74,bib:Levine76,bib:Kozak89-1,bib:Kozak89-2,bib:Ohshima97,bib:Ohshima99,bib:Cuquejo06,bib:Carrique06}
and references therein). This is equivalent, once again, to the
assumption that the properties of a suspension are derived from
those of a single cell placed in a uniform and isotropic host
medium.

The major points of this report are as follows.

First, we outline a new  approach to the effective electrical
properties of nanofluids which is based upon the method of compact
groups of inhomogeneities
\cite{bib:Sushko07,bib:Sushko09,bib:SushkoKr09,bib:Sushko13}.
Within the latter, the nanofluid’s microstructure is modelled in
terms of the complex permittivity profiles of the constitutients.
The desired conductivity and permittivity are expressed in terms
of the moments of these profiles and the volume concentrations of
the constitutients. The moments themselves can be evaluated
without a detailed elaboration of many-particle polarization and
correlation processes in the system. In particular, we present our
theoretical results for a system of hard-core-penetrable-shell
particles embedded into a uniform matrix. The further attention is
focused on the situation where the core radius $R$, the shell
thickness $t$, and the Debye length $r_{\rm D}$ are comparable:
$R/r_{\rm D} \sim t/r_{\rm D} \sim 1$.

Second, we present the results of our experimental studies of the
quasistatic  conductivity and permittivity of $\rm
{Al_2O_3}$-isopropyl alcohol nanofluids as functions of $\rm
{Al_2O_3}$-particle volume concentration $c$. The electrical
parameters of the base liquid (isopropyl alcohol, IPA) samples
were typical of industrial samples, for  the purity requirements
to the base liquid were not mandatory for our purposes. In
contrast, we took advantage of the presence of contaminant ions in
the samples to make simple physical estimates in support of our
model, as shown later.

Finally, we use our theory to process the experimental data
obtained. We show that the observed behavior of $\sigma_{\rm eff}$
and $\varepsilon_{\rm eff}$ can directly be interpreted as a
result of formation of EDLs around the nanoparticles and evaluate
the parameters of these layers.

\section{Theoretical Background}
\label{sec:theor}

\subsection{General Equations}

Suppose that the wavelength $\lambda$ of probing radiation is much
longer than typical distances between the particles dispersed in a
continuous medium. The dielectric response of such a system can
effectively be treated by the method of compact groups of
inhomogeneities \cite{bib:Sushko07,bib:Sushko09}.  These groups
are defined as macroscopic regions within which all the distances
between the particles are much shorter than $\lambda$. With
respect to a probing field with $\lambda \to \infty$, such groups
are actually point-like inhomogeneities, but yet include
sufficiently large numbers $N\gg 1$ of particles to reproduce the
properties of the entire system. Under these conditions, a system
can be viewed as a set of compact groups and characterized by a
certain (modeled according to the system's microstructure) profile
of its complex permittivity. Using methods of the theory of
generalized functions \cite{bib:Vladimirov81} and a special
representation \cite{bib:Weiglhofer89,bib:Sushko04} for the
electromagnetic field propagator, we can, first, extract the
leading compact groups' contributions to the averaged field
$\langle{\rm {\bf{E}}}(\bf{r})\rangle$ and induction $\langle
\bf{D} (\bf{r})\rangle$ in the system without modeling multiple
reemission and short-range correlation effects in depth and,
second, prove that it is these contributions that form the
quasistatic dielectric and conductive properties of the system.
Eventually, the problem reduces to calculating and summing up the
moments of the complex permittivity profile within the system; the
size of compact groups, because of their macroscopicity, turns out
to be an insignificant parameter. In application to suspensions
(see Fig.~1), the essential details of the compact-group approach
are as follows.
\begin{figure}
  \centering
      \includegraphics[width=7.5cm]{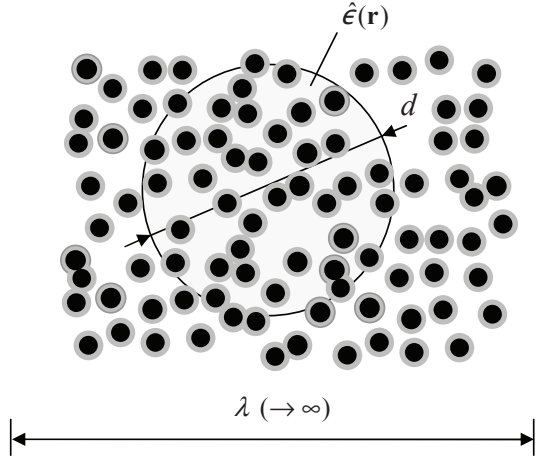}
     \label{Suspension}

\caption{A suspension viewed as a set of macroscopic compact
groups of particles and their EDLs. The number of particles within
a compact group $N\gg 1$, while the linear size of the latter
$d\ll \lambda$. The local value of the complex permittivity in the
suspension is given by (\ref{profile}), where the term $ \delta
\hat{\epsilon}(\bf{r}) $ is due to the presence of a compact group
at point $\bf{r}$.}
\end{figure}

We determine the quasistatic complex permittivity
$\hat{\varepsilon}_{\rm eff}$ of a suspension as the
proportionality coefficient in the relation \cite{bib:Landau84}
\begin{equation} \label{definition}
\langle \bf{D} (\bf{r})\rangle = \langle \epsilon_0 \hat{\epsilon}
(\bf{r}) \bf{E} (\bf{r}) \rangle = \epsilon_0
\hat{\varepsilon}_{\rm eff} \langle \bf{E} (\bf{r}) \rangle
\end{equation}
where
\begin{equation}\label{profile}
\hat{\epsilon}(\bf{r}) = \hat{\varepsilon}_{\rm f} + \delta
\hat{\epsilon}(\bf{r})
\end{equation}
is the local complex permittivity value in the suspension,
$\epsilon_0$ is the electric constant, and the angular brackets
stand for the ensemble averaging or averaging by integration over
the volume.  Expression (\ref{profile}) suggests that the
effective response of the suspension to a probing field is
equivalent to that of an imaginary system made up by embedding the
suspension's constituents, including the base liquid, into a host
having a complex permittivity $\hat{\varepsilon}_{\rm f}$. The
response of this imaginary system is formed by multiple
reemissions and correlations within compact groups of its
constituents (dispersed particles and regions filled with the
suspending liquid). Then the averaged field and induction are
given by
\begin{equation} \label{field} \langle{\rm {\bf  {E}}}\rangle =
{\left[ {1 + {\sum\limits_{s = 1}^{\infty}  {\left( { -
{\frac{{1}}{{3\hat{\varepsilon}_{\rm f} }} }} \right)^{s} \langle
{{\mathop {\left( {\delta \hat{\epsilon} ({\rm {\bf r}})}
\right)^{s}}}}} }}\rangle \right]}\,{\rm {\bf E}}_{0}
\end{equation}
\begin{equation}\label{induction}
\langle{\rm {\bf {D}}}\rangle = \epsilon_0 \hat{\varepsilon}_{\rm
f}{\left[ { 1 - 2 {\sum\limits_{s = 1}^{\infty}  {\left( { -
{\frac{{1}}{{3\hat{\varepsilon}_{\rm f} }}}} \right)^{s}{\langle
{\mathop {\left( {\delta \hat{\epsilon} ({\rm {\bf r}})}
\right)^{s}}}}\rangle} }}  \right]}\,{\rm {\bf E}}_{0}
\end{equation}
where ${\bf E}_0$ is the probing field amplitude in the host of
permittivity $\hat{\varepsilon}_{\rm f}$.

Henceforth, we  take $\hat{\varepsilon}_{\rm f}$ to be equal to
the looked-for permittivity $\hat{\varepsilon}_{\rm eff}$. This
choice implies the use of the Bruggeman-type of electrodynamic
homogenization, which, however, is not identical to  the classical
Bruggeman (mean-field) approach \cite{bib:Bruggeman35}; the latter
considers only the responses of solitary particles to a uniform
effective field \cite{bib:Bergman92}. Then, for a suspension of
identical particles consisting of hard cores (with radius $R$ and
permittivity $\hat{\varepsilon}_1$) and concentric penetrable
shells (with outer radius $R+t$ and permittivity
$\hat{\varepsilon}_2$) and embedded in a suspending liquid (with
permittivity $\hat{\varepsilon}_0$), the moments of $\delta
\hat{\epsilon}(\bf{r})$ are
\begin{equation}\label{moments}
    {\langle
{\mathop {\left( {\delta \hat{\epsilon} ({\rm {\bf r}})}
\right)^{s}}}}\rangle  = (1 - \phi ){(\Delta {\hat{\varepsilon}
_0})^s} +
     c{(\Delta {\hat{\varepsilon} _1})^s}
    + (\phi  - c){(\Delta {\hat{\varepsilon} _{2}})^s}
\end{equation}
where $\Delta \hat{\varepsilon}_i = \hat{\varepsilon}_i -
\hat{\varepsilon}_{\rm eff}$ ($i=1,2,3$), $c$ is the volume
concentration of the hard cores, and $\phi=\phi(c,\delta)$ is that
of the core-shell particles;  $\delta=t/R$ is the relative
thickness of the shells. In view of (\ref{moments}), the series
(\ref{field}) and (\ref{induction}) can be summed up. The
substitution of the sums into (\ref{definition}) gives the
equation for $\hat{\varepsilon}_{\rm eff}$:
\begin{equation} \label{effpermit}
  (1 - \phi )\frac{{{\hat{\varepsilon} _0} - {\hat{\varepsilon}_{\rm eff}}}}{{2{\hat{\varepsilon}_{\rm eff}} + {\hat{\varepsilon} _0}}} +
   c\frac{{{\hat{\varepsilon} _1} - {\hat{\varepsilon}_{\rm eff}}}}{{2{\hat{\varepsilon}_{\rm eff}} + {\hat{\varepsilon} _1}}}
  + (\phi -c)\frac{{{\hat{\varepsilon} _2} - {\hat{\varepsilon}_{\rm eff}}}}{{2{\hat{\varepsilon}_{\rm eff}} + {\hat{\varepsilon}
  _2}}} =0
\end{equation}

Modeling the complex permittivities of the constituents as
$\hat{\varepsilon}_i = \varepsilon_i - {\rm i}{\sigma_i}
/{\epsilon_0 \omega}$ and the effective complex permittivity as
$\hat{\varepsilon}_{\rm eff} = \varepsilon_{\rm eff} - {\rm
i}{\sigma_{\rm eff}} /{\epsilon_0 \omega}$, we can distinguish
different types of the system's response to an external probing
radiation of angular frequency $\omega$. In what follows, we
consider the situation which occurs in the limiting case $\omega
\to 0$. Then (\ref{effpermit}) reduces to two real equations, of
which the governing equation is that for $\sigma_{\rm eff}$:
\begin{equation} \label{effconduct}
  (1 - \phi )\frac{{{\sigma _0} - {\sigma_{\rm eff}}}}{{2{\sigma_{\rm eff}} + {\sigma _0}}} +
   c\frac{{{\sigma _1} - {\sigma_{\rm eff}}}}{{2{\sigma_{\rm eff}} + {\sigma _1}}}
  + (\phi -c)\frac{{{\sigma _2} - {\sigma_{\rm eff}}}}{{2{\sigma_{\rm eff}} + {\sigma
  _2}}}=0
\end{equation}
After it is solved,  $\varepsilon_{\rm eff}$ is found from the
linear equation
\begin{equation} \label{permsustemii} (1 - \phi ) \frac{\varepsilon_{0} \sigma_{\rm eff}-
\varepsilon_{\rm eff} \sigma_{0}}{(2\sigma_{\rm eff} +
\sigma_{0})^2}+ c \frac{\varepsilon_{1} \sigma_{\rm eff}-
\varepsilon_{\rm eff}\sigma_{1} }{(2\sigma_{\rm eff} +
\sigma_{1})^2}  + (\phi - c) \frac{\varepsilon_{2}\sigma_{\rm eff}
- \varepsilon_{\rm eff} \sigma_{2}}{(2\sigma_{\rm eff} +
\sigma_{2})^2} = 0
\end{equation}

The criterion of applicability of (\ref{effconduct}) and
(\ref{permsustemii}) is conveniently represented as
\begin{equation}\label{criterionpractical}
\left|{\sigma_i-\sigma_{\rm eff}}\right|\gg 5.56 \times 10^{-7}
\nu \left|{\varepsilon_i-\varepsilon_{\rm eff}}\right|,\quad i=0,
1, 2,
\end{equation}
where $\sigma_i$ and $\sigma_{\rm eff}$ are expressed in $\mu\rm{
S/cm}$ and $\nu=\omega/2\pi$ is the linear frequency of probing
radiation in $\rm {Hz}$. It follows from the requirement that
$\left| {\rm Im} \left(\hat{\varepsilon}_i -\hat{\varepsilon}_{\rm
eff} \right) \right| \gg \left| {\rm Re} \left(\hat{\varepsilon}_i
-\hat{\varepsilon}_{\rm eff} \right) \right|$, which is expected
to hold true for sufficiently small $\omega$ and is used to derive
(\ref{effconduct}) and (\ref{permsustemii}) from (\ref{effpermit})
within a perturbation scheme with  respect to $\omega$.

The details of the proof of (\ref{moments})--(\ref{permsustemii})
for the case where the averaging procedure is reduced to direct
integration over the region occupied be the suspension can be
found in \cite{bib:Sushko13}; the proof within  the statistical
ensemble averaging will be published elsewhere. It can also be
proven that it is the Bruggeman-type of electrodynamic
homogenization (as described above) that is compatible with the
compact-group approach.

\subsection{Maxwell--Garnett rule}\label{sec:MGrule}

The conductive properties of diluted suspensions are often treated
with the Maxwell--Garnet rule \cite{bib:Maxwell1873,bib:Garnett04}
\begin{equation} \label{MGrule}
\sigma_{\rm eff}=\sigma_0\left(1+2c\cfrac{\sigma_1-\sigma_0}{2
\sigma_0+\sigma_1}\right)\bigg/\left({1-c\cfrac{\sigma_1-\sigma_0}{2
\sigma_0+\sigma_1}}\right)
\end{equation}
To substantiate it, one usually assumes that the dispersed
particles are distributed randomly, their volume concentration is
low, the interaction between them is negligibly small, and only
the contributions from their induced electric dipoles are of
significance. Nonetheless, (\ref{MGrule})  is sometimes used to
describe suspensions with high particle concentrations
\cite{bib:Turner76,bib:Barchini95,bib:Cruz05}, where the mutual
polarization and correlation effects of higher orders come into
play. The effect of the interphase is ignored in deriving
(\ref{MGrule}).

It should be noted that the Maxwell--Garnett rule can be obtained
within the compact-group approach under the suggestions that the
suspending liquid is taken as the host matrix for electrodynamic
homogenization ($\hat{\varepsilon}_{\rm f}=\hat{\varepsilon}_0$)
and the dispersed hard particles have no outer shells.  The
details of relevant calculations can be found in
\cite{bib:Sushko07,bib:SushkoKr09,bib:Sushko09}.

When applying (\ref{MGrule}) in practice, three limiting
representations  of it for $c \ll 1$ can be useful (see, for
instance, \cite{bib:Cruz05}), depending on the relative  magnitude
$X_1\equiv \sigma_1/\sigma_0$ of the particle conductivity, as
compared to the suspending liquid conductivity. Introducing the
relative magnitude $X\equiv \sigma_{\rm eff}/\sigma_0$ of the
suspension conductivity, we have:

(A) If $X_1\ll 1$ (insulating particles), then
\begin{equation} \label{Mginsulating}
X\simeq \frac{1-c}{1+c/2}\approx 1-\frac{3}{2}\, c
\end{equation}

(B) If $X_1\gg 1$ (highly conducting particles), then
\begin{equation} \label{Mgconcucting}
X\simeq \frac{1+2c}{1-c}\approx 1+3\,c
\end{equation}

(C) If $X_1= 1$, then
\begin{equation} \label{Mgequal}
X= 1
\end{equation}

These limiting laws are indeed observed in experiment
\cite{bib:Turner76,bib:Cruz05,bib:Meredith61}, as well as
deviations from them \cite{bib:Meredith61,bib:DeLaRue59}. In
particular, if the volume concentration of the dispersed particles
is increased to values of order 0.05, then, according  to these
rules,  the conductivity of a suspension of insulating particles
is expected to decrease by 8\% and that of highly conducting
particles to increase by no more than 15\%. However, experiment
reveals that in both cases the nanofluid conductivity can increase
by one or even two orders in magnitude (see recent works
\cite{bib:Ganguly09,bib:Posner09,bib:White11,bib:Sikdar11,bib:Konakanchi11,bib:Minea12,bib:Sarojini13}
and references therein). Qualitatively, these deviations are
explained by a complex microstructure of suspensions -- first of
all, by the formation of EDLs around the particles. Since the
Maxwell--Garnett rule lacks any EDL parameters, it fails to give
any quantitative estimates of them and predict their effect on
$X$; neither can it explain a strong enhancement of $\sigma_{\rm
eff}$ with $c$.

\subsection{New Rules For Suspensions of Insulating
Particles }\label{sec:Ourrule}

Now, we use (\ref{effconduct}) to analyze the important case of
suspensions of insulating particles. Despite the fact that this
equation is based upon the Bruggeman-type of electrodynamic
homogenization, it not only reduces, under appropriate
assumptions, to the Maxwell--Garnett limiting laws
(\ref{Mginsulating})--(\ref{Mgequal}), but also can account for
possible deviations from them as functions of the effective
parameters of the interphase.

In (\ref{effconduct}), the interphase is characterized by the
conductivity $\sigma_2$ (or relative conductivity $X_2\equiv
\sigma_2/\sigma_0$, as compared to the suspending liquid
conductivity) and the net volume concentration $\phi -c$. For
spherical hard-core-penetrable-shell particles, their effective
volume concentration $\phi=\phi(c,\delta)$, where $\delta=t/R$ is
the relative thickness of the shell (as compared to the hard-core
radius). The estimates for $\phi(c,\delta)$ are available in
literature. In particular, it has been shown that the analytical
scaled-particle approximation result \cite{bib:Rikvold85}
$$\phi(c,\delta)= 1-
(1 - c)\,e^{-\frac{\left[(1+\delta)^3 - 1\right]c}{1-c}}\,\times$$
\begin{equation} \label{effconc1}
\times  e^{-(1 + \delta)^3 \frac{3 c^2}{2(1 - c)^3} \left[2 -
\frac{3}{1+\delta} + \frac{1}{(1+\delta)^3} - \left(
\frac{3}{1+\delta} - \frac{6}{(1+\delta)^2} +
\frac{3}{(1+\delta)^3}\right) c \right]}
\end{equation}
is in very good agreement with the Monte Carlo simulation results
\cite{bib:Lee88,bib:Rottereau03}
$$\phi(c,\delta)= 1-
(1 - c)\times$$
\begin{equation} \label{effconc}
\times \left\{1-e^{-\left[(1+\delta)^3 - 1\right]c}\,   e^{-(1 +
\delta)^3 \frac{c^2}{2(1 - c)^3} \left[8 - \frac{9}{1+\delta} +
\frac{1}{(1+\delta)^3} - \left(4 + \frac{9}{1+\delta} -
\frac{18}{(1+\delta)^2} + \frac{5}{(1+\delta)^3}\right) c + 2
\left(1 - \frac{1}{(1+\delta)^3}\right) c^2\right]}\right\}
\end{equation}

In what follows, we use estimate (\ref{effconc1}). In the limit of
diluted suspensions ($c\to 0$), all above results take the form
\begin{equation} \label{effconclimit} \phi(c,\delta)
= 1-e^{-(1+\delta)^3c} +O(c^2).
\end{equation}
For hard particles with no shells $\delta=0$ and $\phi(c,0)=c$,
and for core-shell particles $\phi(c,\delta)>c$. If $c\to 0$ and
$\delta \to 0$, then $\phi(c,\delta)-c \approx 3c\delta$. The
dependence of the shell volume concentration $\phi(c,\delta)-c$
upon $c$ for several $\delta$ is shown in Fig.~2.

\begin{figure}[h]
  \centering
      \includegraphics[width=7.5cm]{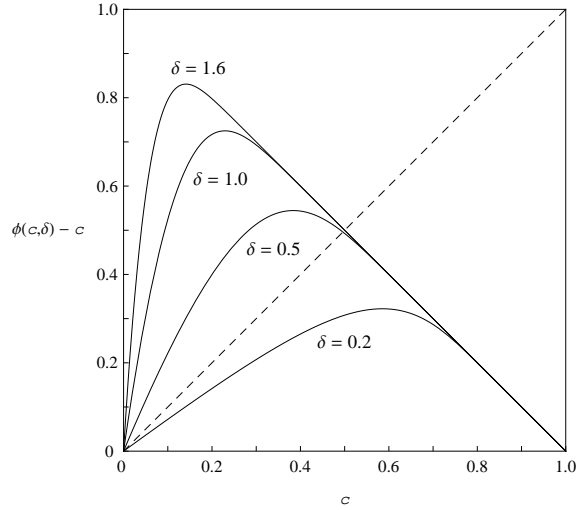}
     \label{EDLVolume}

\caption{ The penetrable-shell volume concentration
$\phi(c,\delta)-c$ as a function of the hard-core volume
concentration $c$ for different $\delta$, according to
(\ref{effconc1}); the dashed line shows the values  of $c$. Only
those values of $\phi(c,\delta)$ and $c$  are physically
realizable for which $c$ does not exceed the close packing value
for the hard cores.}
\end{figure}

For  insulating particles, $X_1 \to 0$, (\ref{effconduct}) reduces
to the quadratic equation
\begin{equation} \label{effconductnanofluid}
  (1 - \phi )\frac{{{1} - {X}}}{{2{X} + {1}}} +
    (\phi -c)\frac{{{X_2} - {X}}}{{2{X} + {X
  _2}}}=\frac{c}{2}
\end{equation}
Simple algebraic manipulations immediately give:

(1) If $X_2\ll X$, then
\begin{equation} \label{Our00}
X\simeq  1-\frac{3}{2}\, \phi
\end{equation}

(2) If $X_2= X$, then
\begin{equation} \label{Our01}
X= 1-\frac{3c/2}{1-(\phi-c)}
\end{equation}

(3) If $X_2\gg X$, then
\begin{equation} \label{Our02}
X\simeq 1+\frac{3(\phi-3c/2)}{1-3(\phi-c)}
\end{equation}

For diluted suspensions with thin particle shells ($c\ll 1$,
$\delta \ll 1$,  $\phi - c \to 0$) and $X\sim 1$, the limiting
rules (\ref{Our00})--(\ref{Our02}) reduce to the Maxwell--Garnet
liming rule (\ref{Mginsulating}). As $\delta$ increases,
deviations from (\ref{Mginsulating}) appear. The greater $X_2$,
the greater they are for a given $c$. As $c$ increases, they also
increase.

As already noted,  the effective conductivity of nanofluids can
increase by up two orders of magnitude as the volume concentration
of embedded insulating nanoparticles increases by a few percent
only. Such behavior is in drastic disagreement with the
Maxwell--Garnett rule (\ref{MGrule}) and its limiting
representations (\ref{Mginsulating})--(\ref{Mgequal}). On the
contrary, rule (\ref{effconductnanofluid}) can potentially explain
this behavior, as seen from its limiting representation
(\ref{Our02}). Two, at least, factors are essential for this: the
presence of sufficiently thick interphase layers between the
particles and the suspending liquid; a relatively high
conductivity of these layers. A positive conductivity increment is
expected for layers with $\delta > \delta_{\rm min} $, where the
threshold value $\delta_{\rm min} $ is, according to
(\ref{Our02}), the solution of the equation $\phi(c,\delta)=3c/2$.
Using (\ref{effconclimit}) and the approximate formula $\ln
(1-3c/2)\simeq -3c/2$, $c\ll 1 $, gives $\delta_{\rm min}\simeq
(3/2)^{1/3}-1\simeq 0.145$. Consequently, the conductivity is
expected to increase with $c$ for suspensions with parameter $R/t<
6.9$. Under the suggestion that $t\approx r_D$, this estimate
agrees  well with the conductivity results \cite{bib:Zukoski85},
especially those derived from the electrophoretic mobility
measurements, for colloids of polystyrene latex particles.

It should be emphasized that electrophoretic measurements deal
with the motion of a practically solitary particle;  the
redistribution of ions near it due to its acquired charge does not
affect the background ionic strength. On the other hand, direct
electrical conductivity measurements deal with averaged properties
of a suspension comprising a small, but finite number of
particles; when charged, these particles can alter the
concentrations of ions in regions both inside and outside the EDLs
(in particular, due to nonspecific adsorption
\cite{bib:Saville83}).

Equation (\ref{effconductnanofluid}) can be used for finding the
effective parameters of suspensions of insulating particles from
low-frequency impedance measurements. Its physically meaningful
solution is
\begin{multline} \label{solutionnanofluid}
\qquad\qquad X =X(c, \delta, X_2) =\\ =\frac{1}{4} \left[2-3 \phi
-X_2- 3c X_2 +3\phi X_2 +\sqrt{\left(2-3 \phi -X_2- 3c X_2 +3\phi
X_2\right)^2+ 4(2-3c) X_2}\right]
\end{multline}
As found above, the asymptotics of $X$ at $X_2 \ll 1$ and $X_2 \gg
1$ are given by  (\ref{Our00}) and (\ref{Our02}), respectively,
with $X_2$ missing.  For diluted suspensions with $\phi \ll 1$,
(\ref{solutionnanofluid}) simplifies to
\begin{equation} \label{solutionnanofluidLimit}
X  = 1 + \frac{3(X_2-1)}{2+X_2}\, (\phi-c) - \frac{3}{2}\,c
+O(\phi^2)
\end{equation}

With formula (\ref{solutionnanofluid}) available, we can  estimate
$X_2$, together with $\delta$, by fitting experimental data for
$X$. An additional opportunity is opened by equation
(\ref{permsustemii}), from which the effective permittivity of
suspensions of insulating particles
\begin{equation} \label{solutionnanofluidpermitivitySimplified}
Y = Y(c, \delta, X_2, Y_1, Y_2) = X\frac{(1-\phi)+ c \cfrac{(2
X+1)^2}{4 X^2}\, Y_1+(\phi-c)\cfrac{(2 X+1)^2}{(2 X+X_2)^2}\,
Y_2}{ (1-\phi)+ (\phi-c)\cfrac{(2 X+1)^2}{(2 X+X_2)^2}\, X_2}
 \end{equation}
where $Y\equiv \varepsilon_{\rm eff} /\varepsilon_0$ and
$Y_i\equiv \varepsilon_i/\varepsilon_0$.  The second term in the
numerator can be neglected (the dependence of $Y$ on $Y_1$ can be
ignored) if $c \ll 1$ and $Y_1 \lesssim 1$.

\section{Experiment}\label{sec:exp}

\subsection{Samples and Measurements' Details}

Two series of nanofluid samples were studied. They were prepared
by the ultrasonic dispersing of $\rm Al_2 O_3$ particles with
passport size of $\rm 50\, nm$ in base liquid (isopropyl alcohol,
IPA) followed by long-term settling and sedimentation of large
particles. The starting mass fraction of $\textrm{Al}_2
\textrm{O}_3$ particles in the stable nanofluids was $\sim 0.1$.
The size control, with dynamic light scattering, revealed that the
dispersed phase consisted of single uncoagulated particles with
approximately equal radii. The samples of different concentrations
were prepared by diluting the initial nanofluid sample with the
base liquid. The mass fraction values for all samples were
determined by the analytical weighing of the net mass of the
$\textrm{Al}_2 \textrm{O}_3$ particles left after the suspending
liquid was evaporated.  The volume concentrations $c$ were
recovered via the particle mass fractions $w$ by
$$
c=\frac{w}{w+(1-w)\rho_1/\rho_0}
$$
where the base liquid density $\rho_0$ and $\rm Al_2O_3$ particle
density $\rho_1$ were taken to be $ \rm 0.785$ and $\rm 3.85\,
g/cm^3$, respectively.

The nanofluid conductivity and permittivity were measured as
functions of  $c$ using an impedance test cell. It consisted of
two rectangular platinum electrodes, with dimensions $\rm 2\, cm
\times 1.5\, cm$ and 3 mm apart, placed in a container with
nanofluid samples, of a $5\,{\rm cm}^3$-volume each. The impedance
data were taken with a Soviet-made automatic digital bridge device
E7-8 at a frequency of $1.00\pm 0.01$ kHz for $c$ ranging from
0.0035 to 0.049 at $20^\textrm{o}$C. This device depends for its
operation on the bridge method with phase-sensitive detectors for
bridge balancing and is designed to measure the electrical
impedance of both two-terminal and multi-branch networks. It also
enables impedance measurements with a bias voltage/current applied
to the sample. We used the operating mode for measuring the
impedance of two-terminal configurations with their equivalent
circuits being the active resistance and capacitance in parallel.
In this mode, the device measurement limits for conductance and
capacitance were 0.1 nS to 1 S and 0.01 pF to 100 $\rm{\mu F}$,
respectively. The instrumental uncertainty was $\pm 0.1 \%$ of
full scale + 1 digit.

It was taken into account that  considerable measurement errors
could be introduced by the polarization at the interface between
the electrodes and the sample, especially when the low-frequency
permittivity of the sample was measured. In  the impedance cell,
the electrode polarization impedance $Z_{\rm p}= R_{\rm p} -{\rm i
}/\omega C_{\rm p}$ was in series with the sample impedance
$Z_{\rm s}= R_{\rm s} -{\rm i }/\omega C_{\rm s}$, so that the
total impedance $Z=Z_{\rm s} + Z_{\rm p}$. We measured $Z$  with
the bridge method; in terms  of an equivalent parallel $R-C$
combination, $Z=R/(1+{\rm i }\omega R C)$ where $R$ and $C$ are
the measured resistance and capacitance, respectively. The use of
platinum electrodes was expected to reduce $Z_{\rm p}$ by several
orders of magnitude. The remaining distorting effect of the
electrode polarization was estimated by the formulas
\cite{bib:Schwan66,bib:Chelidze77}
\begin{equation}\label{experimenttreatment}\frac{R-R_{\rm s}}{R_{\rm s}}=\frac{R_{\rm
p}}{R_{\rm s}}, \quad \frac{C-C_{\rm s}}{C_{\rm s}}=\frac{R_{\rm
p}}{\omega C_{\rm s}R_{\rm s}^2}
\end{equation} Using the latter
one and contrasting our experimental  data for  IPA with
literature data, we found that $R_{\rm p}/R_{\rm s} \approx 0.09$
at most. We believe that  the actual value of $R_{\rm p}/R_{\rm
s}$ was in our experiment several times smaller. Under the
suggestion $R_{\rm s} \approx R$, we recovered with
(\ref{experimenttreatment}) the $R_{\rm s}$ and $C_{\rm s}$ data
and then the conductivities and permittivities of the nanofluid
samples.

The lowest values of  $\sigma_{\rm eff}$ and $\varepsilon_{\rm
eff}$ achieved by dilution of the nanofluid samples were,
respectively, $0.70 \,{\rm \mu S/cm}$ and 37.2 for series I and
$1.05 \,{\rm \mu S/cm}$ and $34.4$ for series II (see Table~1).
These values are higher than those known   for pure IPA, $ 0.06 -
0.2\,{\rm \mu S/cm}$ and $18 - 19$ at room temperatures
\cite{bib:Shell09}, but correlate well with the conductivity value
of $ 1.26 \,{\rm \mu S/cm}$ at $26^\textrm{o}$C, reported in
\cite{bib:Donahue88} for IPA samples recycled after spending. The
conductivity and permittivity of our IPA samples were measured to
be $4.1 \,{\rm \mu S/cm}$ and 28.25, respectively. The former
correlates with data \cite{bib:Handbook06}, according to which the
IPA conductivity is $3.5 \,{\rm \mu S/cm}$ at $25^\textrm{o}$C.

The conductivity and permittivity of $\textrm{Al}_2 \textrm{O}_3$
particles were taken to be $ 1\times 10^{-8}\,{\rm \mu S/cm}$ and
9.2, respectively \cite{bib:Kingery76,bib:Barsoum03}.

\begin{table}
\parbox{16 cm} {{ \small  \textbf{Table 1. Parameters of most diluted nanofluids}}}
\bigskip

\begin{tabular}{|c|c|c|c|}

\hline Series &$c_{\rm m}$&$\sigma_{\rm m}, {\rm
{\mu S/cm}}$ &$\varepsilon_{\rm m}$    \\
\hline
 I, 34 samples& 0.0042   &0.70 &37.2\\
II, 47 samples& 0.0035  &1.05 &34.4\\
\hline
\end{tabular}
\end{table}

\subsection{Experimental Results and Their Processing}

\begin{table}
\parbox{8cm} {{ \small  \textbf{Table 2. Data Fitting Results }}}
\bigskip

\begin{tabular}{|c|c|c|c|c|c|}

\hline series &$\delta$ &$X_2$ &$\delta$ &$X_2$ &$Y_2$    \\
\hline
 I& 1.17 & 35  & 0.90 & 35 &65\\
II& 1.68 & 12  & 1.21 & 12 &12\\
\hline
\end{tabular}
\end{table}

The measurement results for $\sigma_{\rm eff}$ and
$\varepsilon_{\rm eff}$, normalized by their values $\sigma_{\rm
m}$ and $\varepsilon_{\rm m}$ for the most diluted nanofluid
samples with the nanoparticle concentrations $c_{\rm m}$ (see
Table~1), and the fits to these data with the formulas
\begin{equation} \label{fittingK}
K=\frac{X(c,\delta,X_2)}{X(c_{\rm m},\delta,X_2)}
\end{equation}
\begin{equation}
D=\frac{Y(c, \delta, X_2, Y_2)}{Y(c_{\rm m}, \delta, X_2, Y_2)}
\end{equation}
(which enable us to reduce the number of fitting parameters to two
and three, respectively) are represented in Figs.~3 and 4; the
best values of the fitting parameters are summarized in Table~2.
In the next section, we show that most of these values can be
recovered within an elementary model suggesting that the
nanoparticles acquire charge upon dispersing them in base liquid;
this leads to the formation of EDLs around them and alters the
bulk ion concentration in the suspending liquid. The variations in
the relative thickness $\delta$ (at a fixed relative conductivity
$X_2$) of EDLs within each series of the $\sigma_{\rm eff}$ and
$\varepsilon_{\rm eff}$ data may be due to the fact that we ignore
the spatial inhomogeneity of EDLs.

The internal consistency of the fitting results was also verified
with the formula
\begin{equation} \label{inverse}
X_2 = X \frac{3\phi +2(X-1)}{\left(\phi-3c/2
\right)(2X+1)-(1-\phi)(X-1)}
\end{equation}
which follows from (\ref{effconductnanofluid}). Note that
(\ref{inverse}) can be convenient to attack the inverse problem of
recovering $X_2$ and $\delta$ from the experimental $X$ versus $c$
dependence.

\begin{figure}[h]
  \centering
  \begin{subfigure}[b]{0.45\linewidth}
    \includegraphics[width=\textwidth]{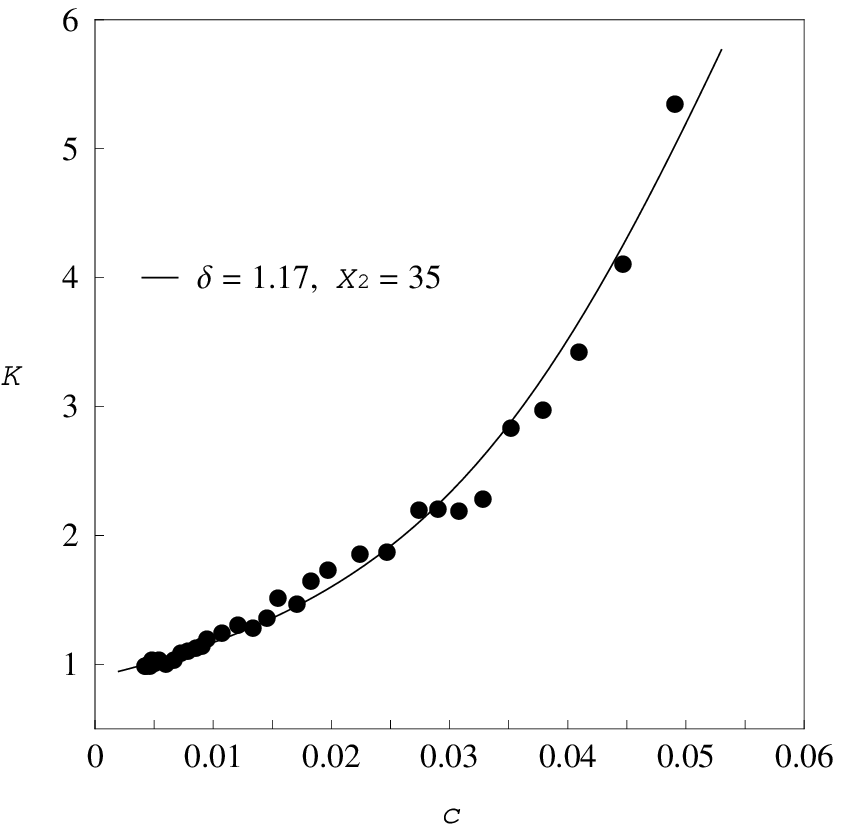}
     \caption{}\label{conductivity1}
     \end{subfigure}
\hfill
  \begin{subfigure}[b]{0.45\linewidth}
      \includegraphics[width=\textwidth]{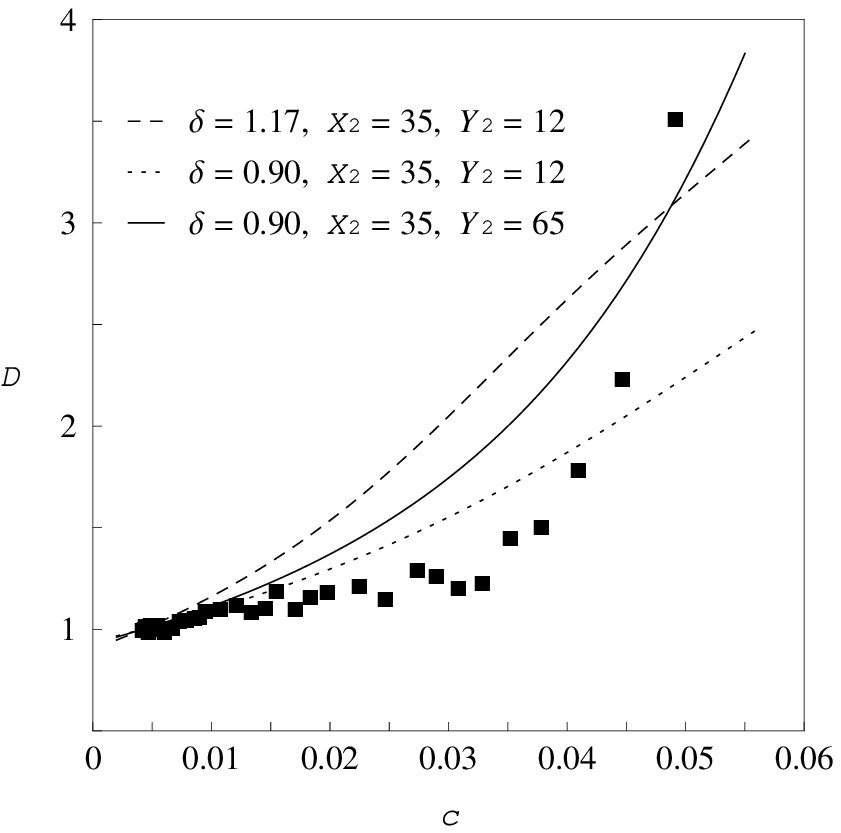}
      \caption{}\label{permittivity1}
 \end{subfigure}

\caption{Fitting the conductivity (a) and permittivity (b) data
for sample series I.}
\end{figure}

\begin{figure}[h]
  \centering
  \begin{subfigure}[b]{0.45\linewidth}
    \includegraphics[width=\textwidth]{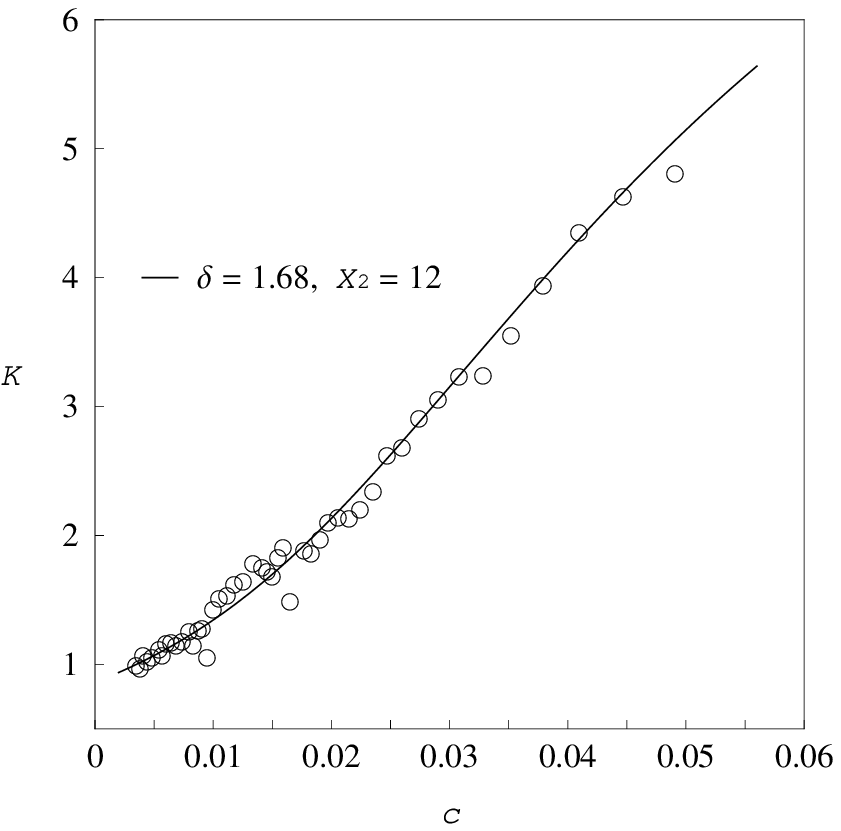}
     \caption{}\label{conductivity2}
     \end{subfigure}
\hfill
  \begin{subfigure}[b]{0.45\linewidth}
      \includegraphics[width=\textwidth]{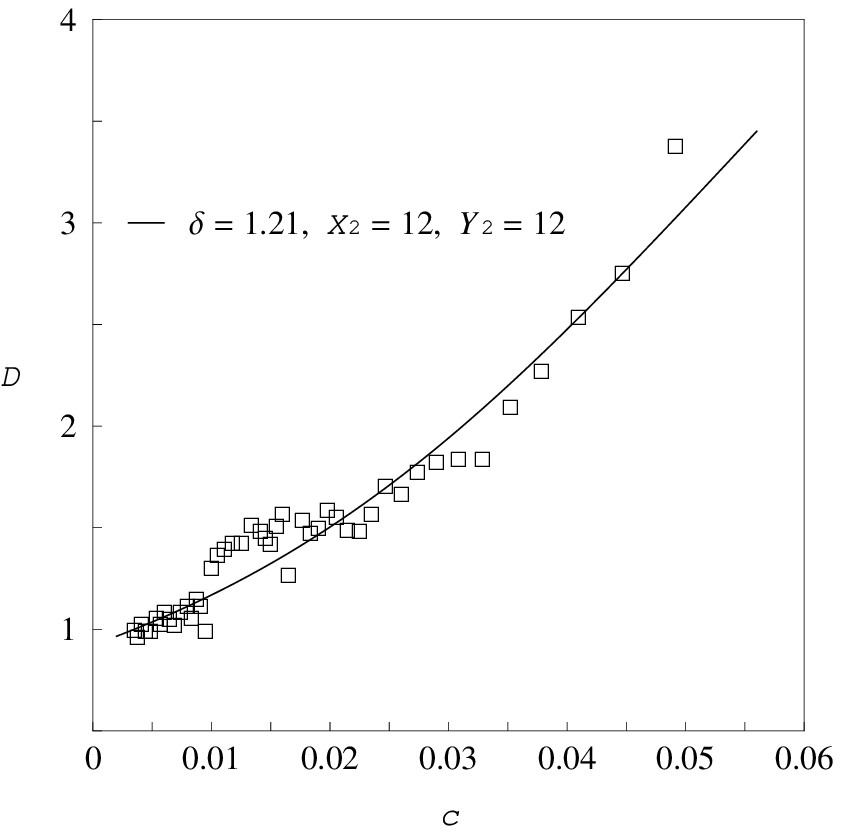}
      \caption{}\label{permittivity2}
 \end{subfigure}

\caption{Fitting the conductivity (a) and permittivity (b) data
for sample series II.}
\end{figure}

\begin{figure}[h]
  \centering
  \begin{subfigure}[b]{0.45\linewidth}
    \includegraphics[width=\textwidth]{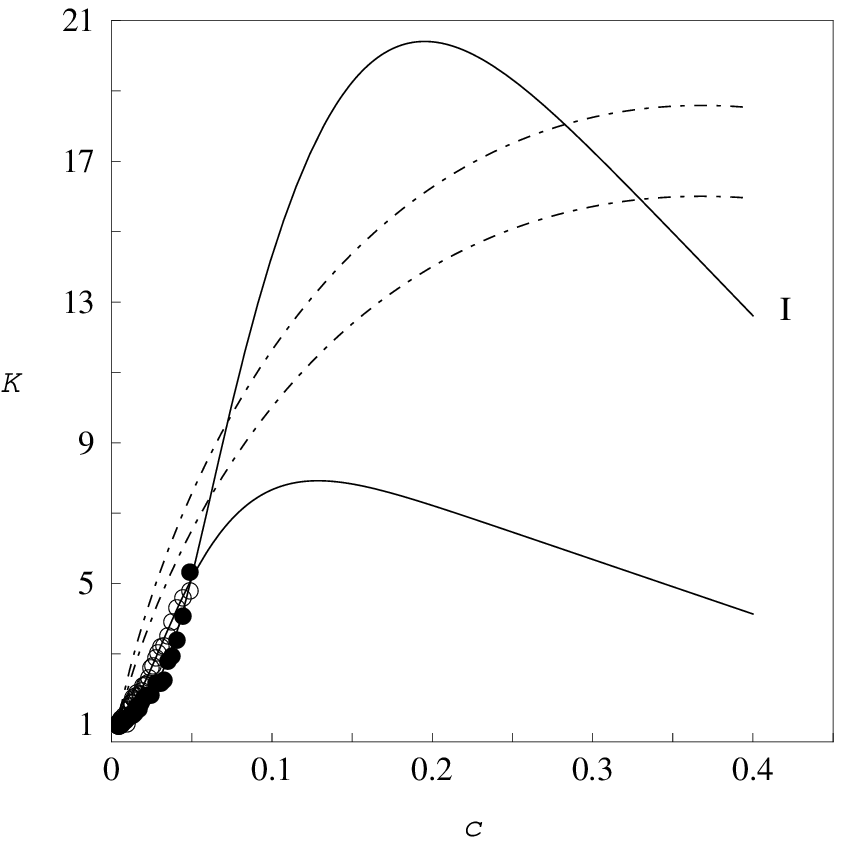}
     \caption{}\label{conductivityextended}
     \end{subfigure}
\hfill
  \begin{subfigure}[b]{0.45\linewidth}
      \includegraphics[width=\textwidth]{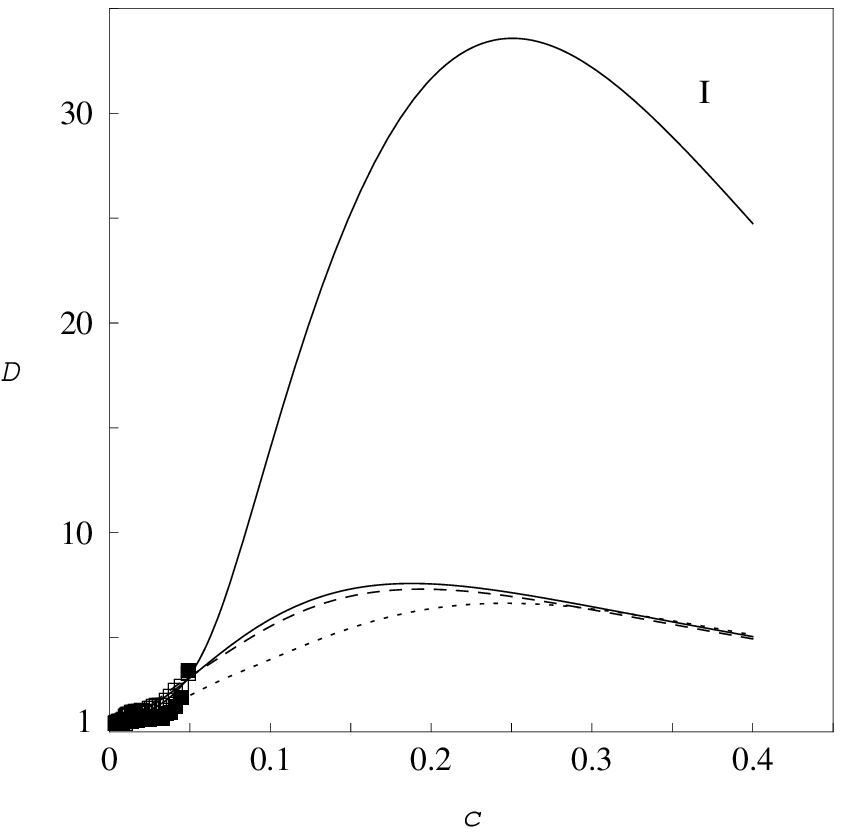}
      \caption{}\label{permittivityextended}
 \end{subfigure}

\caption{ The nanofluid conductivity (a) and permittivity (b) as
functions of $c$, calculated by (\ref{solutionnanofluid}) and
(\ref{solutionnanofluidpermitivitySimplified}) with the parameters
recovered from the low-concentration fits in Figs.~3, 4. The lines
labeled I are the extrapolations of the solid lines in Fig.~3. The
dot-dashed lines in (a) are the graphs of $K^{*}$ versus $c$ at
$c_{\rm m}=0.0042$ (bottom) and 0.0035 (upper).}
\end{figure}

\section {Some  Estimates and Considerations}

Suppose that the conductivity $\sigma_{\rm b}$  of our base liquid
is mainly due to ionic contamination of pure IPA with ions having
unit charges $e$ (of both signs) and equal mobilities $\mu$. The
total concentration of these ions $n_{\rm b}=\sigma_{\rm b}/e
\mu$. Next, assume that after being dispersed into this liquid,
$\rm Al_2O_3$ particles become like-charged, their acquired
charges $Ze$ residing on their surfaces. Via the Coulomb force,
these surface charges cause the formation of concentric diffuse
layers of ions around the particles. Because the total number of
dispersed particles $N_{\rm Al_2O_3}=cV/v$, the total number of
ions on their surfaces $N_{\rm srf}=ZcV/v$; the latter is equal,
in view of the electroneutrality condition, to the total number
$N_{\rm exc}$ of the excess counterions within the diffuse layers.
Here, $V$ is the volume of the nanofluid sample and $v=4\pi R^3/3$
is that of each $\rm Al_2O_3$ particle. The appearance of the
surface- and diffuse-layer-bound ions reduces the concentration of
free ions associated with the suspending liquid to $n_0\approx
(n_{\rm b}V -N_{\rm srf}-N_{\rm exc})/V=n_{\rm b}-2Zc/v$ and,
consequently, the conductivity of this liquid to $\sigma_0=e \mu
n_0\approx\sigma_{\rm b} - 2 e \mu Z c/v$. For sufficiently small
(though finite) $c = c_{\rm m}$, $\sigma_0$ can be equated to the
measured nanofluid conductivity $\sigma_{\rm m}$. Taking
$\sigma_{\rm b}=4.1 \,{\rm \mu S/cm}$, $R= 50\,{\rm nm}$, and $\mu
=1 \times 10^{-7}\,{\rm m^2/(s\cdot V)}$, we obtain the estimate
$Z\approx 1.4\times 10^3$ for the particle charge number in the
most diluted nanofluid of sample series II. Having the measured
permittivity $\varepsilon_{\rm m}$ of this nanofluid, we further
find that the Debye length $r_{\rm D} \approx 85\, {\rm nm}$ and
the ratio $R/r_{\rm D} \approx 0.58$. The average concentration of
the excess counterions in each diffuse layer $n_{\rm exc
}=Z/v_{\rm lr}$ where $v_{\rm lr}
=\frac{4\pi}{3}\left[(R+t)^3-R^3\right]$ is the volume of the
layer. The relative conductivity $X_2\equiv \sigma_2/\sigma_0$ of
the layers can be evaluated as $X_2 =1+ Z/(v_{\rm lr}n_0)$. For
$t= r_{\rm D}$, we obtain  $\delta\approx 1.7$ and $X_2 \approx
23$. In the case of sample series I, $Z\approx 1.3 \times 10^3$,
$r_{\rm D} \approx 109\, {\rm nm}$ and $R/r_{\rm D} \approx 0.46$.
For $t= 0.75\,r_{\rm D}$, we find that $\delta\approx 1.6$ and
$X_2 \approx 34$. Lastly, we note that $n_{\rm exc} \approx 1.4
\times 10^{23} \,{\rm m}^{-3}$ for both sample series; this is
considerably greater than  $n_0 \approx 4.4 \times 10^{21}\, {\rm
m}^{-3}$ (series I) and $n_0 \approx 6.6 \times 10^{21}\, {\rm
m}^{-3}$ (series II). Evaluating $\sigma_2$ as $\sigma_2 \approx
n_{\rm exc} \Lambda^0/N_{\rm A}$ ($\Lambda^0$ is the limiting
ionic conductivity and $N_{\rm A}$ is Avogadro's number) and
taking for $\Lambda^0$ typical literature values of $10$ and $5\,
{\rm mS}\cdot {\rm m}^2/{\rm mol}$, we obtain $\sigma_2\approx 23
$ and $12\, {\rm \mu S}/{\rm cm}$, respectively. These  are very
close to the experimentally-recovered values, given by $X_2
\sigma_0$. The estimates by $\sigma_0 \approx n_0 \Lambda^0/N_{\rm
A}$ and with the same $\Lambda^0$'s give, rather satisfactorily,
$\sigma_0 \approx 0.73 $ (series I) and $ 0.55\, {\rm \mu S}/{\rm
cm}$ (series II).

Despite their simplicity, the above estimates are internally
consistent and agree sufficiently well with our data processing
results.  This fact is a strong argument for the model being
proposed. In this respect, several more points are worth noting.

First, our experimental curves in Figs.~3, 4 exhibit pronounced
curvatures. Provided $R/r_{\rm D} < 6$, similar behavior is
revealed by the $\sigma_{\rm eff}$ versus $c$ curves for latex
suspensions which were first exhaustively ion-exchanged and then
dialyzed against deionized water for 24 to 48 h
\cite{bib:Zukoski85}; no curvature was observed in latex
suspensions which had been cleaned extensively by ion exchange. As
pointed out in \cite{bib:Zukoski85}, these facts indicate a high
sensitivity of $\sigma_{\rm eff}$ to ionic contamination.
Quantitative cell-model-based estimates \cite{bib:Carrique09} for
the effect of atmospheric $\rm {CO_2}$ contamination on
$\sigma_{\rm eff}$ of aqueous salt-free suspensions lead to
similar conclusions. In view of \cite{bib:Carrique09}, the
formation of a plateau-like segment on the experimental
$\varepsilon_{\rm eff}$ versus $c$ dependence in Fig.~3~b, where
our theory and experiment deviate most, may result from a
higher-than-expected level of uncontrolled contamination for
sample series I. If so, the variations in the numerical values of
the fitting parameters for the two sample series are reasonably
explicable.

Second, assuming that the particles acquire charge by adsorbing
contaminant ions,   we can explain the initial decrease of
$\sigma_{\rm eff}$ at very small $c$ by depletion of the base
liquid in those ions.  As $c$ increases further, the particles'
conducting EDLs may begin to  form percolation-type paths. This
effect (most likely, enhanced by others, such as dissociation of
base liquid, Donnan exclusion \cite{bib:Lyklema95} etc.) should
cause $\sigma_{\rm eff}$ and $\varepsilon_{\rm eff}$ to start
increasing with $c$. The conductivity increment sign reversal has
indeed been observed in suspensions with low ionic strength (see,
for instance, \cite{bib:Zukoski85,bib:Sarojini13}).

Third, as the graphs in Fig.5 reveal, our model predicts that at
higher $c$, when the overlapping of EDLs comes into play, the rate
of growth, $r_{\sigma}$, of $\sigma_{\rm eff}$ and that of
$\varepsilon_{\rm eff}$ may decrease, reach zero levels, and even
become negative. Such trends have been observed for $\sigma_{\rm
eff}$, but at lower $c$ (see, for instance,
\cite{bib:Posner09,bib:White11,bib:Sikdar11,bib:Zawrah15}). For a
salt-free medium, the decrease of $r_\sigma$ with $c$ can be
explained \cite{bib:White11} by counterion condensation
\cite{bib:Alexander84,bib:Ohshima02,bib:Ohshima03} near the
particles which develop high surface charge due to the protonation
or deprotonation of surface groups. In terms of our model, this
mechanism can be accounted for through, at least,  the reduction
of the effective thickness of EDLs without changing the
conductivity outside. On the other hand, if the charging of
particles occurs mainly through preferential adsorption of
contaminant ions, then the bulk ion concentration should decrease
with $c$ and the effective $\delta$ should increase. Consequently,
the indicated scenario for $r_{\sigma}$ is expected at lower
values of $c$, compared to those shown in Fig.~5 (see also
Fig.~2).

Finally, in view of our estimate $n_{\rm exc }\gg n_0$, it might
be suggested that the $c$-behavior of $\sigma_{\rm eff}$ in our
model should resemble that predicted by the electrokinetic theory
\cite{bib:Ohshima02,bib:Ohshima03} for dilute suspensions of
spherical particles in a salt-free medium containing counterions
only. Based upon Kuwabara's cell model and arguments
\cite{bib:Imai52}, Ohshima showed that if the total surface charge
of the particle exceeds a critical value given, for univalent
counterions, by
$$Q_{\rm crit}= Z_{\rm crit} e= 4\pi \varepsilon_{\rm r}\epsilon_0 R\left(\frac{k_{\rm B}T}{e}\right) \ln \left(1/c\right)$$
where $\varepsilon_{\rm r}$  is the relative permittivity of the
medium, $k_{\rm B}$  is the Boltzmann constant, and $T$  is the
absolute temperature, then the counterion concentration takes a
practically constant value
$$n^{*}=   \frac{3 \varepsilon_{\rm r}\epsilon_0 k_{\rm B}T}{e^2R^2} c\,\ln \left(1/c\right)$$
except in the region very near the particle surface.
Correspondingly \cite{bib:Ohshima02,bib:White11}, the suspension
conductivity is
$$\sigma^{*}_{\rm eff}\approx \frac{n^{*}}{N_{\rm A}} \Lambda^0=
\frac{3 \varepsilon_{\rm r}\epsilon_0 k_{\rm B}T}{N_{\rm A}
e^2R^2}\Lambda^0 c\,\ln \left(1/c\right)$$ and parameter (25)
becomes
$$K^{*}=\frac{c\,\ln \left(1/c\right)}{c_{\rm m}\ln \left(1/c_{\rm m}\right)}$$
Taking $\varepsilon_{\rm r}\approx \varepsilon_{\rm m}$, $c\approx
c_{\rm m}$ and the previous values for the other parameters, for
$Z=(1.3 - 1.4)\times 10^3$ we find $Z_{\rm crit} \approx 171 -
179$ and, consequently, $n^{*} \approx 1.4 \times 10^{21} \, {\rm
m}^{-3}$, $\sigma^{*}_{\rm eff}\approx  0.24\,{\rm \mu S/cm}$
(series I) and $n^{*} \approx 1.1 \times 10^{21} \, {\rm m}^{-3}$,
$\sigma^{*}_{\rm eff}\approx  0.10\,{\rm \mu S/cm}$ (series II).
As expected, these values are considerably smaller than those for
our samples.  Simultaneously,  $K^{*}$ as a function of $c$
reveals an increase of the same order as $K$ does (see Fig. 5 a).
It is also seen that the initial segments of these graphs have
different curvatures. In our opinion, this fact is indicative of
different nature of the models considered: the cell model is a
mean-field approach, while ours is a percolation-type theory, with
the effective microstructure parameters $\delta$, $X_2$ and $Y_2$.

\section{Conclusion}
\label{sec:discussion}

The main results of this report can be summarized as follows. The
general expressions for the quasistatic effective electrical
conductivity and permittivity of systems of
hard-core-penetrable-shell particles, obtained earlier with the
method of compact groups of inhomogeneities, have been scrutinized
to formulate new mixing  rules for suspensions of charged
insulating particles. These rules express the conductivity and
permittivity of a suspension in terms of the effective geometric
and dielectric parameters of its constituents (the particles,
their EDLs, and the suspending liquid). They effectively
incorporate higher-order correlations and polarizations in the
suspension without in-depth modeling these processes. Thus the
problem of electrodynamic homogenization of a suspension is
reduced to a simpler problem of finding the effective parameters
of its constituents; the latter can be attacked with, say, the
standard electrokinetic theory. On the other hand, the rules
obtained can be used to extract the effective parameters of the
constituents from low-frequency impedance measurements.

Such measurements have been done for two series of ${\rm
Al_2O_3}$-isopropyl alcohol nanofluids to obtain their
conductivity and permittivity as functions of $\rm
{Al_2O_3}$-particle volume concentration. The functional form of
the new rules is sufficient to fit these dependences even under
simplest suggestions about the mechanism of charging of the
particles and its impact on the system. In particular, the
recovered values of the parameters agree well with elementary
theoretical estimates based on the idea that the particles acquire
charges through preferential adsorption of contaminant ions. The
possible effects of other mechanisms on the effective conductivity
and permittivity of suspensions are also discussed.

We hope that our results will stimulate further studies in the
field, including the performance of specially-designed
experiments.

\section*{Acknowledgements}

This research was supported in part by Ministry of Education and
Science of Ukraine through Grant 0113U001377. We also thank
Prof.~D.K.~Das and Dr. S.~Sikdar for providing copies of their
articles.

\bigskip

\end{document}